\documentclass[12pt,a4paper,leqno]{article}
\usepackage{graphicx}%
\title{
H{\"o}lder-exponent-MFDFA-based test for long-range correlations in pseudorandom sequences}
\author{Nikolay K. Vitanov$^{1,2, \footnote{corresponding author},
\footnote{email:vitanov \@ mpipks-dresden.mpg.de}}$, Khristo Tarnev$^{1,3}$ and Holger Kantz$^{1}$}
\date{$^{1}$
Max Planck Institute for the Physics of Complex Systems, N{\"o}thnitzer Str. 38,
01187, Dresden, Germany \\
$^{2}$
Institute of Mechanics, BAS, Akad. G. Bonchev Str., Block 4,
1113 Sofia, Bulgaria\\
$^{3}$
Department of Applied Physics, Technical University of Sofia,1000 Sofia, Bulgaria
}

\begin{document}
\maketitle
\begin{abstract}
We discuss the problem for detecting long-range correlations in sequences
of values obtained by generators of pseudo-random numbers. The basic idea is that
the H{\"o}lder exponent  for a
sufficiently long sequence of uncorrelated random numbers has the fixed value
$h=1/2$. The presence of long-range correlations leads to deviation from this value. We
calculate the H{\"o}lder exponent by the method of multifractal detrended
fluctuation analysis (MFDFA). We discuss  frequently used tests for
randomness, finite sample properties of the MFDFA, and the conditions for a correct
application of the method. We observe that the fluctuation function $F_{q}$ used
in the MFDFA reacts to trends
caused by low periodicity presented in the pseudo-random number generator.
In order to select appropriate generators from the numerous programs 
 we propose a test for the ensemble
properties of the generated pseudo-random sequences  with respect to their
robustness against presence of
long-range correlations, and a selection rule which
orders the generators that
pass the test. Selecting  generators that successfully pass the ensemble test and have good
performance with respect to the selection rule is not enough. For the selected generator
we have to choose appropriate pseudo-random sequences
for the length of the sequence required by the solved problem.
This choice is based on the
closeness of the H{\"o}lder exponent of the generated sequence to its value $1/2$
characteristic for the case of absence of correlations.
\end{abstract}

\maketitle
\section{Pseudorandom sequences, H{\"o}lder exponent and multifractal detrended fluctuation analysis}
Computer random number generators have many applications in physics as
for an example in the
Monte Carlo methods \cite{wong97,fishmann96} or in the nonlinear time series
analysis \cite{k1,k2}.
Computers implement deterministic algorithms, hence
they are not able to generate sequences of truly random numbers. Thus the
randomness of the sequences is relative: What can be random enough for one
application may not be random enough for another application. The computer
generated sequences are called
pseudorandom, and in order to be  appropriate for scientific applications they
have to satisfy many requirements and must pass extensive statistical tests.
For an example the
period of the generator (the number of different results before the generator repeats itself)
must be as large as possible. Another requirement is that the generated numbers should not
be correlated among themselves. A significant class of such correlations are
the long-range correlations which decay much slower than exponentially with
time or distance and are observed in many systems in the Nature \cite{st1,st2}.
The autocorrelation function $c_{n}=\langle \eta_{i} \eta_{i+n} \rangle$
of a pseudorandom sequence $\{ \eta_{\alpha} \}$ ideally should be a Kronecker $\delta_{n,0}$.
$c_{n}$ is not a convenient tool for
detecting long-range correlations among the numbers in the sequence since the presence or
absence of weak correlation at large $n$ is usually masked by
statistical fluctuations. Hence we have to apply a separate test to detect
them. In this paper we propose such a test which is
based on the simple fact that the H{\"o}lder exponent
for the sequence of uncorrelated random numbers has the fixed value of $1/2$
(for more details see the Appendix). The deviation from this value for
large $n$ is evidence for possible long-range correlations in the
sequences generated by the used pseudo-random numbers generator.
\par
The H{\"o}lder exponent is a measure of irregularity of a curve at a singular point
\cite{mandelbr,tricot}. Let us consider a function $z(t)$. Its $\tau$-oscillation
at value $t$ is
\begin{equation}\label{def1}
{\rm osc}_{\tau} (t)= \sup_{\mid t-t' \mid \le \tau} \hskip.1cm z(t') -
\inf_{\mid t-t' \mid \le \tau} \hskip.1cm z(t').
\end{equation}
 The graph of $z(t)$ is fractal
if
${\rm osc}_{\tau}(t)/\tau \to +\infty$
uniformly with respect to $t$. For differentiable $z(t)$ the ratio
${\rm osc}(t) /(2 \tau)$ tends to
$dz/dt$ for $\tau \to 0$. When the limes superior of the ratio is infinite, there is no
derivative, and the H{\"o}lder exponent measures the singularity of the graph
of $z(t)$ in this point. A function $z(t)$ is H{\"o}lderian of exponent $h$ if a 
constant $c$ exists such that for all $t'$
\begin{equation}\label{def2}
\mid z(t) - z(t') \mid \le c \mid t- t' \mid^{h}
\end{equation}
 or in terms
of $\tau-$oscillations ${\rm osc}_{\tau}(t) \le c \tau^{h}$. If ${\rm osc}_{\tau}(t) \ge c
\tau^{h}$ the function $z(t)$ is anti-H{\"o}lderian of exponent $h$ at the point $t$.
The relation between  $c_{n}$
and the H{\"o}lder exponent for the case of Hurst noise is discussed in the Appendix.
We recall from there
the relationship (for large $n$)
\begin{equation}\label{recall}
c_{n}=h(2h-1)n^{2h-2}.
\end{equation}
For the case of pure white noise ($c_{n}=0$), $h=1/2$.
When $h >1/2$, $c_{n}>0$, i.e., the series of data are correlated and when
$h<1/2$, $c_{n}<0$, i.e., the data series are anti-correlated.
%\begin{figure*}[t]
%\vskip-1cm
%\includegraphics[angle=0,height=4.0cm,width=4.5cm]{figx1a.eps} \hskip.2cm
%\includegraphics[angle=0,height=4.0cm,width=4.5cm]{figx1b.eps} \hskip.2cm
%\includegraphics[angle=0,height=4.0cm,width=4.5cm]{figx1c.eps}
%\caption{Influence of the length of sequence on the $h(q)$ spectrum. Generator {\sf ran0},
%ensemble averages for ensemble of $25$ time series.
%The segment length $s$ is between $s_{min}=10$ and $s_{max}=1000$ for all panels. The
%lengths of sequences are $10^{4}$ points for panel (a) $10^{5}$ points for
%panel (b) and $10^{6}$ points for panel (c).
%The $h(q)$ spectra are denoted by solid lines
%and the dashed lines denote the correspondent $h(q)$ spectrum plus/minus the standard
%deviation of the mean for the correspondent value of $q$.}
%\end{figure*}
\par
We shall calculate the H{\"o}lder exponent for our pseudorandom sequences by means of the
MFDFA method \cite{kant02} which is briefly described in the Appendix.
We perform all calculations by means of
the MFDFA(1),i.e., linear fit of the trend in  each segment (When the
fit is made by polynomial of $p$-th order the method is denoted as MFDFA(p)). Our choice $p=1$
respects the fact that no trend is to be expected in the investigated
sequences of numbers. Below we shall discuss a test for
long-range correlations and not for a trend in the generated sequences of
pseudorandom numbers. If such a trend exists it has to be detected by other
tests. But even in the case when these additional tests are not performed
the MFDFA reacts to the presence of trends and nonstationarities. This
sensitivity is know to exist for DFA \cite{ph1,ph2} and as we shall observe below
it exists for the case of MFDFA too.
\par
We use as random number generators several generators from \cite{press} namely
 {\sf ran0},  {\sf ran2}, {\sf ran3}, a quick and dirty generator
which
we call {\sf qdg}  as well as the generator based on
the program G05CAF from the NAG library.
{\sf ran0} is the minimal standard generator which has to be satisfactory
for most applications but if its parameters are not appropriately chosen
the generator can have correlations. The generator {\sf ran2} is claimed
to be very good one and without serial correlations up to limits of its
floating point precision. {\sf ran3} is based on the algorithm due to Knuth
\cite{knuth}. {\sf qdg} is based on a cycle containing the lines of code \\
\begin{flushleft}
$\dots$\\
jran=mod(jran*ia+ic,im) \\
ran=float(jran)/float(im)\\
$\dots$
\end{flushleft}
where im, ia, ic are appropriately chosen parameters.
The above generators are not
tested and compared with respect of long-range correlations.
\par
The paper is organized as follows. In the following section we discuss the properties
of the investigated generators with respect for standard tests for randomness. Finite sample
properties of the MFDFA method are investigated in section III in order to select
interval of appropriate values
for the parameter $q$ of the method for which the test for
randomness has to be performed. In the
section IV the behavior of the fluctuation function of the MFDFA method is discussed
for the investigated generators and it is shown that the fluctuation function is a
good indicator for deviation from fractality and presence of trends in the generators.
In section V we discuss an ensemble test for long-range correlations in the
random number generators as well as rules for selection of (i) most appropriate
generator for the required length of the sequence and (ii) most appropriate sequences
generated by a selected generator.
Some concluding remarks are summarized in the last section where we discuss MFDFA(0) and
MFDFA(1) with respect to their sensitivity to detect deviations from the case 
$h=1/2$. In the
Appendix we describe shortly the relationship between the correlation function and the
H{\"o}lder exponent as well as the MFDFA method.
\section{Behavior of generators with respect to standard tests}
In order  to be sure that {\sf ran0}, {\sf ran2}, {\sf ran3}, { \sf qdg} and
{\sf G05CAF}
produce sequences close to random ones we  have tested them by standard tests such as
a frequency test or  a serial correlations test \cite{wong97}. The general difficulty with
statistical test on finite sequences lies in the statistical fluctuations. For the latter
tests they are perfectly known.
We expect that the generators generate sequences of pseudorandom
numbers with
histograms uniform in some  interval (
$[0,1]$ for {\sf ran0}, {\sf ran2}, {\sf ran3}, {\sf G05CAF} and
$[-1,1]$ for {\sf qdg}).  In order to perform the frequency test we sort a generated  sequence
of $N$ numbers into $B$ bins with expected mean value of
the numbers in each bin  $\overline{M}=N/B$. If the actual number of random numbers
in the $j-$th bin is $M_{j}$ we can construct the quantity
\begin{equation}
\chi^{2}_{\nu} = (1/\nu)\chi^{2} = (1/\nu) \{\sum_{j=1}^{B}[(M_{j} - \overline{M})^{2}
/\overline{M}]\}
\end{equation}
with $\nu = B-1$ degrees of freedom. $\chi^{2}_{\nu}$ must be different from $0$ because of
presence of some fluctuations in a histogram of a finite sequence of numbers. However
the value  of $\chi_{\nu}^{2}$
must be not too large because large values are the evidence for the
concentration of numbers in some bins and thus the generator is not random.
$\chi^{2}$
was calculated for $\nu=49$ degrees of freedom and for a good random number generator
$\chi^{2}$ has the value of about $40$ for a sequence of 50000 pseudorandom numbers.
%
%\begin{table} [h]
%\caption{\label{tab:table1} $\chi^{2}$ test for several pseudorandom number
%generators. {\sf idum}=2 for {\sf ran0, ran2, ran3}. im=6075, ia=106, ic=1283
%for {\sf qdg}. }
%\begin{ruledtabular}
%\begin{tabular}{cccccc}
%Number of points & {\sf ran0} & {\sf ran2} & {\sf ran3} & {\sf qdg} & {\sf G05CAF} \\
%\hline
%$10^{4}$ & 46.30 & 39.83 & 45.13 & 7.14 & 42.25 \\
%\hline
%$10^{5}$ & 46.29 & 44.38 & 57.13 & 2.35 & 41.34 \\
%\hline
%$10^{6}$ & 42.04 & 43.29 & 38.84 & 16.54& 59.68 \\
%\hline
%$10^{7}$ & 41.25 & 57.33 & 41.38 & 165.84 & 56.35 \\
%\end{tabular}
%\end{ruledtabular}
%\end{table}
%
Characteristic values for the investigated generators are
presented in Table I. We note (i) the good performance of the
quick and dirty generator
{\sf qdg} for small length of the generated pseudorandom sequences
opposite to the case of large length of the sequence
 and  (ii) the fact
that the different generators have different performance for
different lengths of the
sequence.  We can conclude that all generators except the {\sf qdg}
passed this test and we observe that the choice of the
appropriate generator depends on the lengths of the pseudorandom sequence
we need.
\par
As a second test we have calculated the autocorrelation at lag $\nu$
\begin{equation}
c_{\nu}=\frac{1}{\sigma^{2}} \langle (x_{n}-\langle x \rangle)(x_{n-\nu}-
\langle x \rangle) \rangle
\end{equation}
where $\langle x \rangle$ and $\sigma^{2}$ are the mean and the variance
of the sequence. Several typical results for the investigated
generators are presented in Table II.
%\begin{table}
%\caption{\label{tab:table2} Autocorrelations $c_{\nu}$ for several
%pseudorandom number generators. Length of generated sequences is $10^{5}$
%values. {\sf idum}=2 for {\sf ran0, ran2, ran3}. im=6075, ia=106, ic=1283
%for {\sf qdg}. }
%\begin{ruledtabular}
%\begin{tabular}{cccccc}
%$\nu$ & {\sf ran0} & {\sf ran2} & {\sf ran3} & {\sf qdg} & {\sf G05CAF}  \\
%\hline
%$10^{2}$ & -0.0061  & -0.00098 & -0.0013 & 0.0017  & 0.00086 \\
%\hline
%$10^{3}$ & -0.0008 & -0.0033 & -0.0076 & -0.00009 & 0.0018 \\
%\hline
%$10^{4}$ & -0.0029 & 0.0043 & -0.0061 & -0.000098& 0.0055 \\
%\hline
%$5 \cdot10^{4}$ & -0.0045 & 0.0013 & -0.0023 & -0.0048 & 0.0019 \\
%\end{tabular}
%\end{ruledtabular}
%\end{table}
The small  values of the autocorrelations show that the generators
successfully pass this test. This  fact  supports the observation
that the long-range correlations could be quite good masked so that the
pseudorandom number generators with such correlations could pass the standard
correlation tests.
We can conclude that except for the cases of very inappropriate generators
the standard tests do not supply us with much information about
the question if the pseudorandom generator we want to use is free from long-range correlations.
We have to design other tests to detect such correlations. The test and rules which are
discussed below are based on the use of the H{\"o}lder exponent which is calculated by
means of the multifractal detrended fluctuation analysis (MFDFA) method.
\section{Finite sample properties of MFDFA}
Before we apply the MFDFA method to our pseudorandom number
sequences we have to
understand its finite sample properties. For  infinite uncorrelated
sequences of values the H{\"o}lder exponent
has the fixed value $h=1/2$. We discuss here in detail finite sample properties of sequences
generated by two of the generators: the generator {\sf ran2} which
is claimed to be quite good in \cite{press}, and the generator {\sf ran0}.
The investigation of sequences generated by {\sf C05CAF} and {\sf ran3} leads
to the same conclusions. We
estimate the H{\"o}lder exponent $h(q)$ from a sample of
length $N$ for ensemble of pseudorandom sequences (For the role of
the parameter $q$ see the Appendix ).
The MFDFA method will be consistent when
$h(q) \to 1/2$ with increasing length $N$ of the sample, and unbiased when the ensemble average
$\langle h_{(N)}(q) \rangle \to 1/2$.
Our observation is that the MFDFA method is consistent but biased
for large $\mid q \mid$. For the investigation of the finite sample properties of the two
generators we have  used at least 10 ensembles each of $25$ random sequences (i.e.  at least 250
different sequences).
For larger number of sequences in the ensembles the results for $h(q)$
do not change significantly
and the only significant effect is the decreasing of the standard
deviation of the mean.
We test the consistency of the method by keeping parameters unchanged except for
the length of the time series which is increased . Several characteristic results
for the generator {\sf ran0} are presented in Fig. 1.
These results as well the results from other similar calculations starting
from different seeds of the generator parameter {\sf idum} of {\sf ran2}
show that the MFDFA method is consistent.
However, the method is biased for large $\mid q \mid$ as it can be seen
from Fig. 1 and Fig. 2.
%\begin{figure*}[t]
%\includegraphics[angle=0,height=4.0cm,width=4.5cm]{figx2a.eps} \hskip.2cm
%\includegraphics[angle=0,height=4.0cm,width=4.5cm]{figx2b.eps} \hskip.2cm
%\includegraphics[angle=0,height=4.0cm,width=4.5cm]{figx2c.eps}
%\caption{Influence of the length of segment $s$ on the $h(q)$ spectrum of the generator
%{\sf ran2}. The investigated sequences of numbers are obtained for {\sf idum=2} and have length
%of $250 000$ values. The
%$h(q)$ spectrum is denoted by a solid line. The two doted lines denote the $h(q)$ spectrum
%plus/minus the standard deviation of the mean for the correspondent value of $q$.
%Panel (a): $h(q)$ spectrum for $s$ between $s_{min}=10$ and $s_{max}=1000$. Panel (b):
%$h(q)$ spectrum for $s$ between $s_{min}=100$ and $s_{max}=10000$. Panel (c):
%$h(q)$ spectrum for $s$ between $s_{min}=1000$ and $s_{max}=25000$.}
%\end{figure*}
 We observe that for
small $\mid q \mid$ (the numerical experiments lead us to the conclusion that small is
$ \mid q \mid \le 2$)  $h(q)$ stays around $0.5$ but for larger $\mid q\mid$ significant
deviations exist (i) when we increase the length of the sequence and keep $s_{mix}$ and
$s_{max}$ constant (Fig.1), and (ii) when we keep the length of the sequence constant and
increase $s_{max}$ (Fig. 2). A part of the bias of the method for large $\mid q \mid$ can be
removed by increasing sample size because for small sample size we have limited number of
values with low probability. Nevertheless our advice is that all test for long-range correlations
must  be performed for small $\mid q \mid$.
As generators generate only pseudorandom sequences we can not expect to obtain
exactly $1/2$ as a value for $h$. Our numerical investigation has shown that
 differences of $0.005$ are admissible, i.e., the generator (from the point of view of
ensembles of generated sequences)  can be considered safe with respect to
long range correlations when for small $\mid q \mid$ (between $0$ and $2$)
its ensemble $h(q)$ spectrum  is between $0.495$ and $0.505$.
\section{Behavior of the fluctuation function}
When we investigate sequences from a  pseudo-random numbers generator first of all
we have to check if the
fluctuation function $F_{q}(s)$ from the MFDFA method really
scales as a power of $s$ for different values of $q$.  We expect
$F_{q}(s)$ to be a straight line on the log-log scale (panels (a),(b),(c)) of Fig.3) .
Deviations from this behavior are evidence for
a problem. In panel (d)  we see a broken line with a break point approximately at $1/2$ of the
period of generator which is characteristic for sequences with
periodic trend.  Hence the
MFDFA method reacts on trends in the generated sequences i.e. it  can be used as
a warning message for presence of trends too.
In addition, even if
the fluctuation function is a straight line, but the
value of $h$ is significantly different from $0.5$ for small $\mid q \mid$,
this is an evidence for presence of long-range correlations in the sequence of
pseudo-random numbers.
In panels (a)-(c) of Fig. 3 we observe that with increasing length of the sequence
for {\sf ran2} $F_{q}(s)$ comes closer to a straight line. This could be expected because the
larger length leads to a better statistics as the form of the  sequence histogram
approaches the ideal assumed histogram form. 
Panel (f) shows a comparison between fluctuation functions for 
characteristic sequences generated by {\sf ran0}
and {\sf ran2}. We observe that
{\sf ran2} is more robust
than {\sf ran0} with respect to long-range correlations because the
fluctuation function lines for different $q$ are closer to straight lines for the
case of the generator {\sf ran2} for all range of segment lengths $s$. 
Fluctuation functions for {\sf ran0} are more dispersed for large lengths of the
segment $s$.

\par
We haven't obtained satisfactory results for the generator {\sf qdg}.
This means that the fluctuation function does not scale as power law of the
length of the segment (as in the case of the vales of im=$6075$, ia=$106$,
ic=$1283$), or the scaling exponent is quite significant from $0.5$ as for an example
in the case of the values of parameters im=$233280$, ia=$9301$, ic=$49297$.
\section{A test and  selection rules}
On the basis of obtained results we can formulate a test and rules of selection among
pseudo-random numbers generators and among sequences generated by a generator.
The test determines the conditions under which a
random number generator can be considered to generate large amount of sequences
 free from long-range correlations. We note here that no
absolute test exists, i.e.,  a generator can be good for one required length of
the sequence but another generator could be better for another length of the sequence.
Let us have several
pseudo-random numbers generators and  let us want to choose these of them which generate
large amount of sequences free of long-range correlations. This choice can be made on the
basis of the following test on the ensembles of pseudorandom sequences.
\par
{\sf For a given random number generator take at least 10 different ensembles each of  at least 25
pseudorandom sequences and calculate
the H{\"o}lder exponent by means of MFDFA(1) method for  $\mid q \mid \le 2$. If for all
ensembles the
fluctuation function $F_{q}(s)$ scales as power law for all $q$
and the H{\"o}lder exponent is between
$0.495$ and $0.505$ the random number generator can be considered to be able to
generate large amount of sequences free from long-range correlations.}
\par
Generators which pass the above test
have to be preferred with respect to generators that fail the test. Thus one
can select generators with small probability of generating sequences possessing
long-range correlations (for the required length of the sequence). The next question
which arises is how to order the appropriate generators and to select one of them.
The answer is given by a selection rule which is based on
a power-law scaling property of the fluctuation function and on closeness of corresponding
H{\"o}lder exponent to $1/2$ and states
\par
{\sf  Let us have two pseudo-random numbers generators which pass the above test. Let us
calculate the fluctuation function
$F_{q}(s)$ for at least 10 different ensembles  each of at least $25$
sequences for the two generators.
The generator which
has closer to power law form of $F_{q}(s)$ for all $q$ and
for which the $h(q)$ is closer to $1/2$ is more robust with respect to
long-range correlations}.
\par
An extensive investigation leaded us to the following ranking of the generators with
respect of the test and the selection rule
\begin{enumerate}
\item
{\sf ran2}
\item
{\sf G05CAF}
\item
{\sf ran3}
\item
{\sf ran0}
\end{enumerate}
\par
Despite the fact that a generator passes the test and it is selected
by the selection rule each
pseudorandom sequence has to be tested separately i.e. the generator can be chosen among
a manifold of generators   but nevertheless some  of its sequences can
have long-range correlations among their values. Therefore
after choosing the most appropriate generator (for the required  by the solved
problem length of the pseudorandom
sequence) we have to test every generated sequence for presence of long range correlations.
The best sequences are selected by a selection rule which is analogous to the above
selection rule namely that the appropriate sequences
have H{\"o}lder exponents most close to $1/2$ for $\mid q \mid <2$.
Several results for pseudorandom sequences from different generators are presented in
Table III. The sequences are chosen for illustration of
the fact that each generator can generate sequences for which $h$ is considerably different
from $1/2$.
\section{Concluding remarks}
In this paper we have used
 the multifractal detrended fluctuation analysis  to investigate the behavior of the
H{\"o}lder exponent for sequences of pseudorandom numbers obtained by several
 random number generators. Theoretically
the H{\"o}lder exponent for a large enough sequence of random numbers must have a fixed value
$h=1/2$ regardless of the order $q$ of the fluctuation function  in the MFDFA method.
The deviations from randomness lead to three kinds of changes:
(i) The fluctuation
function $F_{q}(s)$
is not a straight line on a log-log plot. This is  evidence for presence of some trend in the
generated sequences i.e. the generator is very bad one;
(ii) When $F_{q}(s)$ scales as a power law of $s$, $h(q)$
is a straight line
significantly different from $1/2$ for all values of $q$  for $\mid q \mid  \le 2$.
This indicates presence of long-range correlations but if $h$ do not depend on $q$ the generated sequence has monofractal
properties up to smallest investigated length of the segments of the MFDFA;
and (iii) $h(q)$ could be close
to $1/2$ for some values of $\mid q \mid$ but not for all  values for $ \mid q \mid$
for $\mid q \mid \le 2$. This means presence of long-range correlations and
multifractal properties of the generated sequences.
The existence of bias at large $\mid q \mid$
means that MFDFA must be used very carefully when one calculates
characteristic  fractal quantities for sequences of values with multifractal properties.
Our case here is a monofractal one (theoretically $h$ has a
value, independent on $q$)  but nevertheless we have to take into  account
this bias by restriction on the values of $\mid q \mid$.
\par
The simplest variants of the MFDFA from the point of view of the fitting
polynomial are MFDFA(0) and MFDFA(1). Above we have used MFDFA(1).
It is possible to use MFDFA(0) in the formulated test and selection rules (i.e.
to use MFDFA without local detrending).
Let us discuss the properties of MFDFA(0) and
MFDFA(1) with respect to their application to sequences of pseudorandom numbers.
Let us assume that our sequence of pseudorandom numbers is divided into $N_{s}$ segments
each of length $s$ and let us write the profile function $Y$ for the $\nu$-th segment
as
\begin{equation}\label{comp1}
Y[(\nu-1)s+i] = i^{1/2} + \delta_{\nu}(i)
\end{equation}
where $\delta_{\nu}(i)$ is the deviation of $Y$ from $i^{1/2}$ at the $i$-th value of
the $\nu$-th segment. For the fitting polynomial we assume
\begin{equation}\label{comp2}
y_{\nu}(i)=a_{\nu}i+b_{\nu}
\end{equation}
where $a_{\nu}$ and $b_{\nu}$ are constant coefficients. If $a_{\nu}=b_{\nu}=0$ we
have MFDFA(0) variant of the MFDFA method. When $a_{\nu} \ne 0$ and (or) $b_{\nu} \ne 0$
we have the MFDFA(1) variant of the MFDFA method.

From (\ref{comp1}) and (\ref{comp2}) we obtain for the variation $F^{2}(\nu,s)$ for the
$\nu$-th segment (for large enough values of $s$)

\begin{eqnarray}
F^{2}(\nu,s) = \frac{s}{2} - a_{\nu} b_{\nu} s + \frac{1}{s} \sum_{i=1}^{s}
\delta_{\nu}^{2}(i)+ \frac{1}{s} \sum_{i=1}^{s} 2 i^{1/2} \delta_{\nu}(i) - \nonumber \\
\frac{2 a_{\nu}}{s} \sum_{i=1}^{s} i \delta_{\nu}(i) - \frac{2 b_{\nu}}{s}
\sum_{i=1}^{s} \delta_{\nu}(i) + \frac{a_{\nu}^2 s^{2}}{2} + b_{\nu}^{2}  - \nonumber \\
a_{\nu} s^{3/2} -b_{\nu} s^{1/2} \nonumber \\
\end{eqnarray}
Let for simplicity below $q=2$. Then for the fluctuation function we have
\begin{equation}\label{comp3}
F_{2}(s) = \sqrt{P+Q+S}
\end{equation}
 where
\begin{equation}\label{comp4}
P=\frac{1}{2 N_{s}} \sum_{\nu=1}^{2 N_{s}} \left( \frac{s}{2} - a_{\nu} b_{\nu} s \right)
\end{equation}
contains the terms corresponding to the ideal case,

\begin{equation}\label{comp4}
Q=Q^{*} + Q^{**}
\end{equation}
where
\begin{equation}\label{qst1}
Q^{*}= \frac{1}{2 s N_{s}} \sum_{i=1}^{s} \sum_{\nu=1}^{2 N_{s}} [\delta_{\nu}^{2}(i) +
2 i^{1/2} \delta_{\nu}(i)]
\end{equation}

\begin{equation}\label{qst2}
Q^{**}=\frac{1}{2 s N_{s}} \sum_{i=1}^{2 N_{s}} \sum_{\nu=1}^{2 N_{s}}
[2 a_{\nu} i \delta_{\nu} (i) - 2 b_{\nu} \delta_{\nu} (i)]
\end{equation}
are the terms containing the fluctuations $\delta_{\nu}(i)$. $Q^{*}$ depends
only on the fluctuations and $Q^{**}$ depends on the fluctuations and on the trend.
Finally
\begin{equation}\label{comp4}
R= \frac{1}{2N_{s}} \sum_{\nu=1}^{2 N_{s}} \left( \frac{a_{\nu} s^{2}}{2} +
b_{\nu}^{2} - a{\nu}^{3/2} - b_{\nu} s^{1/2}\right)
\end{equation}
contains the terms which depend only on the local trend.
\par
For the case of MFDFA(0) ($a_{\nu}=b_{\nu}=0$) $P=Q^{**}=0$. In the ideal case
$Q^{*}$ should be negligible and $F_{2}(2) \propto s^{1/2}$ for large $s$ i.e.
the H{\"o}lder exponent is $h=1/2$. In the real case $Q^{*}$ could lead to
deviation from $h=1/2$ and if these deviations are large this is an indicator for
presence of problems in the generated sequence.
\par
For the case of MFDFA(1) the sensitivity can be higher as we have more terms that
can affect the value of the H{\"o}lder exponent. The term $P$ in this case is again proportional to
$s^{1/2}$ independent on the values of the local trend coefficients $a_{\nu}$ and
$b_{\nu}$. For large enough $s$ and large enough sequence of random numbers all
other terms should be negligible and $h \propto 1/2$. In the real situation deviations
can come from $Q^{*}$, $Q^{**}$ and from R which for large $s$ is proportional to $s^{2}$.
When the local trends are correlated (i.e. some kind of global trend is presented) then
$R$ ( which for large $s$ could become larger than $P$) could lead to $h$ close to $1$ instead
to $h \approx 1/2$ as in the case without long-range correlations. Similar situation
is observed for other values of $q$. For an example when $q=4$ the term $P$ is  a
sum of terms of the kind $s^{2}(1/4 + 2 a_{\nu}^{2} b_{\nu}^{2} + a_{\nu} b_{\nu})$
(i.e. $P \propto s^{2}$) and the dominant for large $s$ member of $R$ is proportional to
$s^{4}$. In summary if we do not want to use MFDFA with local detrending we can
base our test and rules on MFDFA(0). In this case the deviation from the ideal case
(i.e. from $h=1/2$) is evaluated on the basis of quantities like $Q^{*}$. When we use
MFDFA(1) we can gain additional sensitivity with respect to deviations of $h$ from
$1/2$.

\par
Finally we note that the MFDFA is not the only possibility for calculation
of H{\"o}lder exponent. Another method is the WTMM (wavelet transform
modulus maxima) method \cite{arn1,arn2,arn3,arn4,arn5}. MFDFA is shown to
have slight advantages for negative $q$ and short time series \cite{kant02}.
For long time series the WTMM could have advantages with respect to MFDFA and thus
for length of sequences larger than $10^{7}$ values the formulated tests above should
be used on the basis of the results of the WTMM method. In such a case instead of
scaling of the fluctuation function $F_{q}(s)$ one has to study the scaling of the
partition function $Z_{q}(a)$ used in the WTMM.
\section{Acknowledgments}
N.K.V. thanks to Alexander von Humboldt Foundation and to NFSR of the Ministry of Education and
Science of Bulgaria for support
of his research through the Grant \# MM 1201/02.
\begin{appendix}
\section{The importance of the exponent $h$ and MFDFA}
Let us consider the general one-dimensional random walk in discrete time.
It is a sum of steps which can be either discrete or continuous. Let the
walking particle starts at the origin. After $N$ steps its position $X_{N}$
is a sum of $N$ mutually independent random variables. We are interested in
a situation when these variables have the same distribution function $F$
of mean $\mu$ and finite variance $\sigma^{2}$. It can be shown on the
basis of the central limit theorem that for large $N$ $X_{N}$ is approximately
normally distributed with mean $N \mu$ and variance $N \sigma^{2}$
\cite{sinai,cox,weiss}. Thus the standard deviation of this random walk scales
as $N^{h}$ with $h=1/2$.
\par
In order to understand better the H{\"o}lder exponent $h$
let us consider a sequence of observations $\eta_{i}, i=1,2,,\dots,N$. We choose a reference
size $m_{0}$ and sizes $m$ such that $mp = N$, where $p$ is the number of segments each of
size $m$. For fixed $m$ we calculate the mean value  $M(m)$
 and the standard deviation
$S(m)$ for each segment and construct the relative dispersion $R(m)=S(m)/M(m)$.
From general manifold of all possible data sequences we shall
consider these for which we can observe
\begin{equation}\label{r1}
[R(m)/R(m_{0})] = (m/m_{0})^{h-1},
\end{equation}
where for simplicity we consider $h$ to be a constant. Mandelbrot
\cite{mandelbr} calls similar fluctuations and noise Hurst noise in order
to notify the important contribution of Hurst \cite{hurst} to the
research of processes with long-range correlations.
By means of (\ref{r1}) we obtain
a system of equations for the
correlation functions
\begin{equation}
c_{\tau}=  \langle \eta_{i} \eta_{i+\tau} \rangle = C_{1}/C_{2}
\end{equation}
where
\begin{equation}
C_{1}= [N/(N-\tau)] \sum_{i=1}^{N-\tau}
\eta_{i} \eta_{i+\tau} - (\sum_{i=1}^{N} \eta_{i})^{2}/N
\end{equation}
\begin{equation}
C_{2}= \sum_{i=1}^{N} \eta_{i}^{2}-
(\sum_{i=1}^{N
}
\eta_{i})^{2}/N,
\end{equation}
for $\tau=1,2,\dots,n$. The system is
\begin{equation}
\sum_{i=1}^{n-1} (n-1) c_{i} = (1/2) \left(n^{2h}-n \right)
\end{equation}
and it has the solution \cite{bass1,bass2}
\begin{equation}
c_{n}=(1/2) n^{2h} \left[ \left(1+ (1/n) \right)^{2h} - 2 + \left(1-(1/n)
\right)^{2h} \right].
\end{equation}
For very large value of $n$ we can represent the term in $[\dots]$ as Taylor series and as a result
we
obtain the relationship between the autocorrelation and the H{\"o}lder exponent
\begin{equation}\label{relation}
c_{n}=h(2h-1)n^{2h-2}.
\end{equation}
\par
Recently a multifractal detrended fluctuation analysis method has been proposed
for the analysis of long-range correlation of nonstationary time series \cite{kant02}.
Here we present the variant of the method useful for calculation of significantly different
from zero positive H{\"o}lder exponents.
The first step of the method is to calculate the mean
$\langle x \rangle$ of the investigated time series.
Then we calculate the profile function
$Y_{i}=\sum_{k=1}^{i} (x_{k} - \langle x \rangle), \hskip.5cm i=1,2,\dots,N$.
After calculation of $Y_{i}$
we divide the time series into segments and  calculate the variation for
each segment. The division is into $N_{s}=$int$(N/s)$ segments and because the obtained segments would
not include some data at the end of the investigated time series,  additional $N_{s}$ segments
are added, which start from the last value of the sequence in the direction to the first
value of sequence.  In order to calculate the variation  we have to calculate the
local trend (the fitting polynomial $y_{\nu}(i)$ for each segment of length $s$
where $s$ is between an appropriate minimum and  maximum value).
The variations are defined as
\begin{equation}
F^{2} (\nu,s) = \frac{1}{s} \sum_{i=1}^{s} \left \{ Y[(\nu-1) s+i] -y_{\nu}(i)
\right \}^{2}
\end{equation}
for the first $N_{s}$ segments  and
\begin{equation}
F^{2} (\nu,s) = \frac{1}{s} \sum_{i=1}^{s} \left \{ Y[N-(\nu-N) s +i] -
y_{\nu}(i) \right \}^{2}
\end{equation}
for the second $N_{s}$ segments.
Finally we construct the $q$-th order fluctuation function
\begin{equation}
F_{q}(s) =  \{ [1/(2N_{s} )] \sum_{\nu=1}^{2N_{s}} [ F^{2} (\nu,s)]^{q/2}
\}^{1/q}.
\end{equation}
For
monofractal time series $F_{q}(s)$ has to scale as $s$ of constant power $h$
which for sequences of random numbers has the value $1/2$. Even in presence
of local correlations extending up to a characteristic range $s^{*}$ the
exponent $h=1/2$ would be unchanged when $s>>s^{*}$. If the correlations do
not have characteristic lengths the exponent $h$ would be different from $1/2$
\cite{bunde,rangar}.
\end{appendix}
\bibliography{vitanov}
\bibliographystyle{unsrt}
%
%\begin{thebibliography}{99}
%\bibitem{wong97}
%        S. S. M. Wong. {\sl Computational Methods in Physics and Engineering}. World
%        Scientific, Singapore (1997).
%\bibitem{fishmann96}
%        G. S. Fishmann. {\sl Monte Garlo. Concepts, Algorithms and Applications}. Springer,
%        New York (1996).
%\bibitem{mandelbr}
%        B. B. Mandelbrot. The Fractal Geometry of Nature. Freeman, New York, 1982.
%\bibitem{tricot}
%        C. Tricot. Curves and Fractal Dimension. Springer, New-York, 1995.
%\bibitem{press}
%        W. H. Press, S. A. Teukolsky, W. T. Vetterling, B. P. Flannery. Numerical
%        Recipes in Fortran. The art of Scientific Computing. Cambridge University Press,
%        Cambridge, 1992.
%\bibitem{bass1}
%        J. B. Bassingthwaighte, R. P. Beyer. 
%        Physica D, {\bf 53}, 71-84 (1991)
%\bibitem{bass2}
%        J. B. Bassingthwaighte, L. S. Liebovitch, B. J. West. Fractal physiology. Oxford
%        University Press, New York, 1994.
%\bibitem{kant02}
%        J. W. Kantelhardt, S. A. Zschiegner, E. Koscielny-Bunde, S. Havlin, A. Bunde,
%        H. Eugene Stanley. 
%         Physica A {\bf 316}, 87-114 (2002).
%\end{thebibliography}

\newpage

\begin{figure}[t]
\includegraphics[angle=0,height=4.0cm,width=4.5cm]{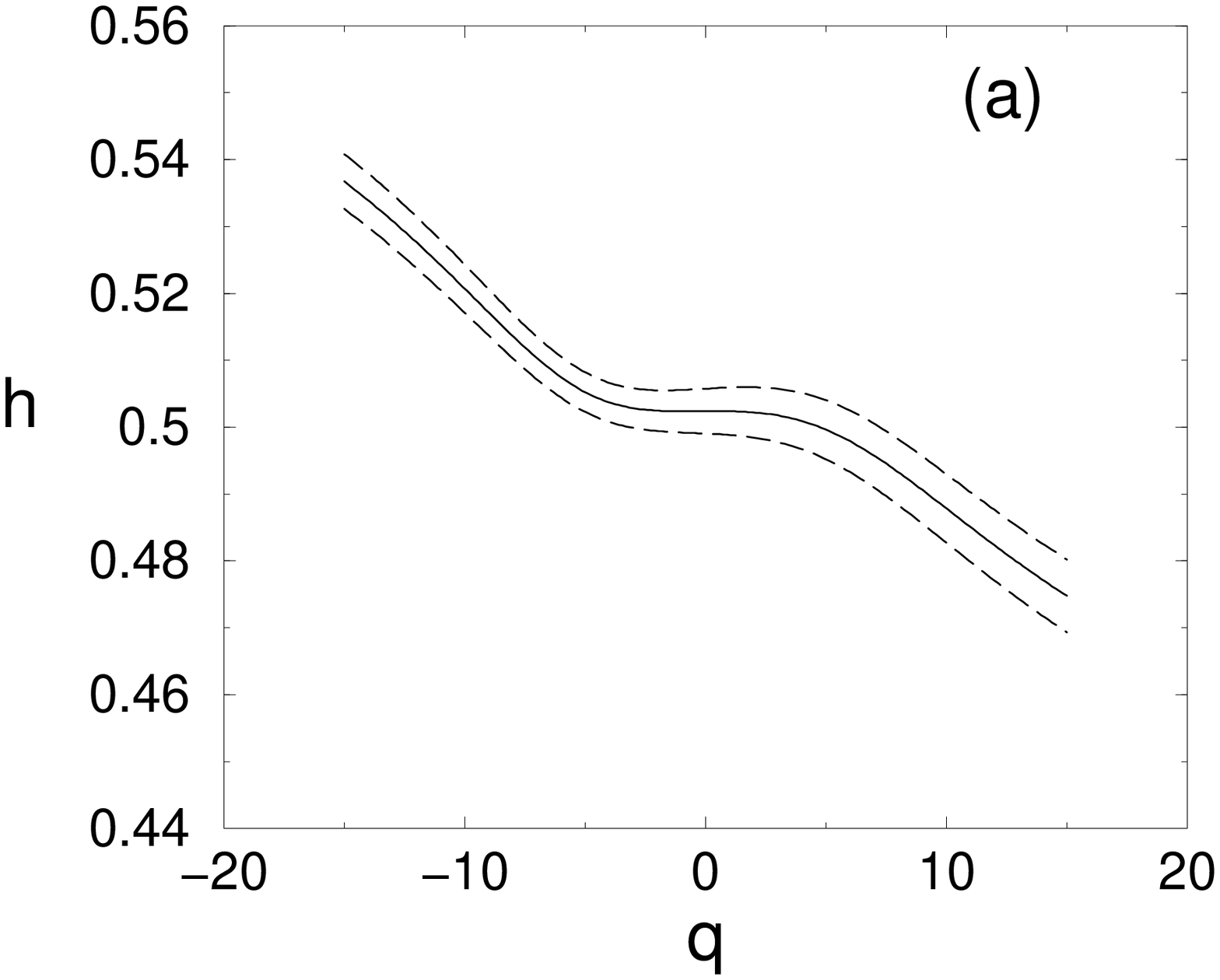} \\
\includegraphics[angle=0,height=4.0cm,width=4.5cm]{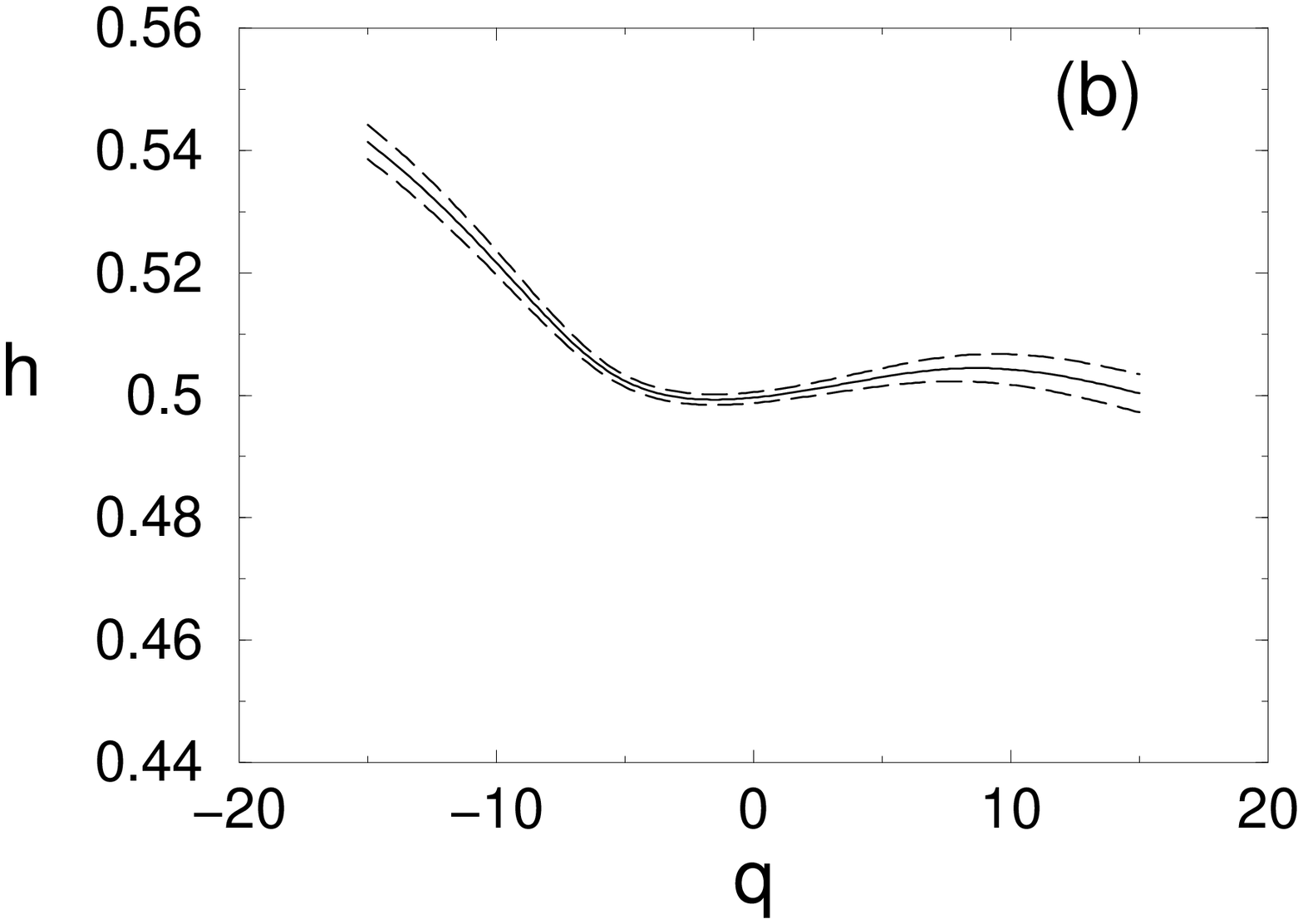} \\
\includegraphics[angle=0,height=4.0cm,width=4.5cm]{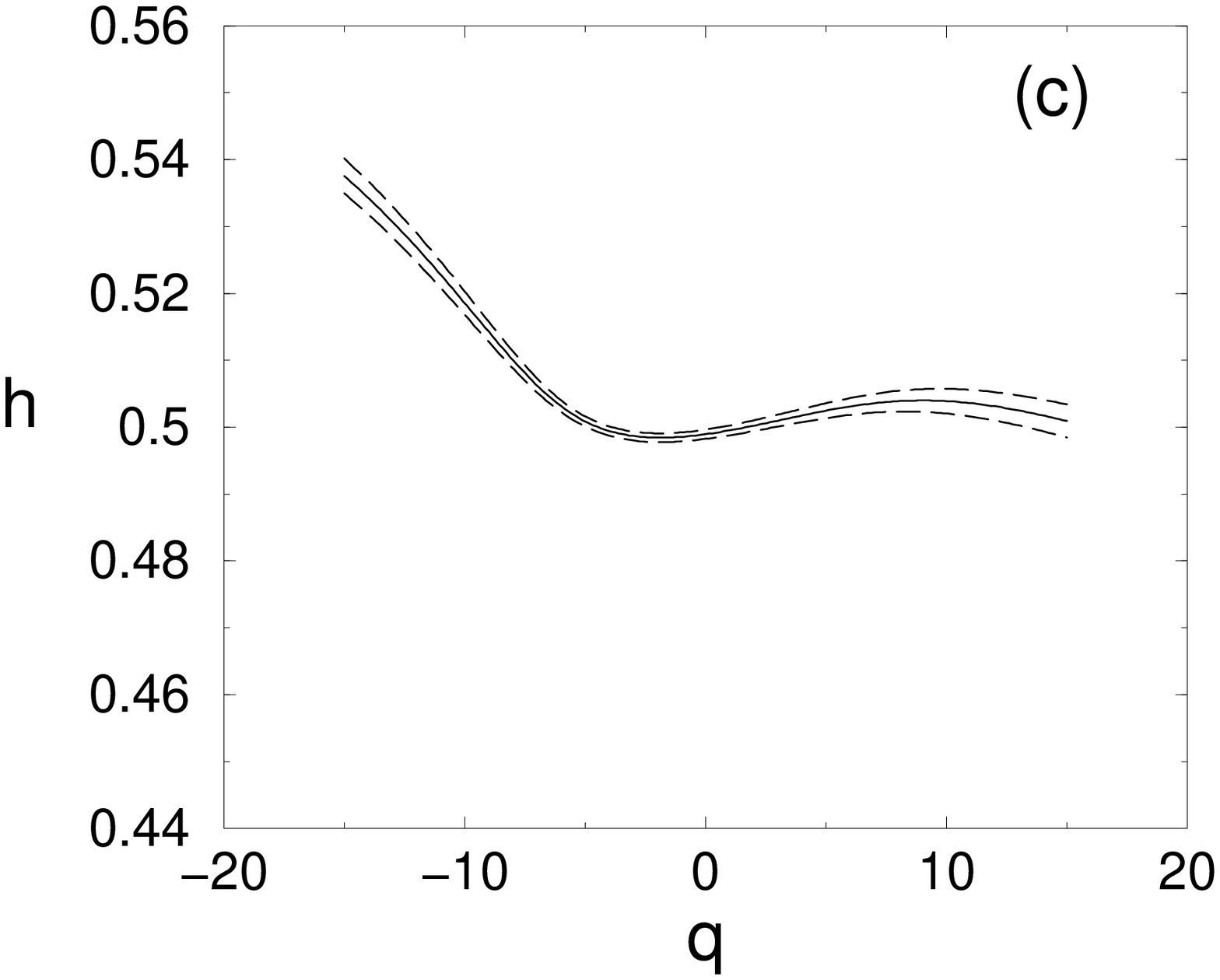}
\caption{Influence of the length of sequence on the $h(q)$ spectrum. Generator {\sf ran0},
ensemble averages for ensemble of $25$ time series.
The segment length $s$ is between $s_{min}=10$ and $s_{max}=1000$ for all panels. The
lengths of sequences are $10^{4}$ points for panel (a) $10^{5}$ points for
panel (b) and $10^{6}$ points for panel (c).
The $h(q)$ spectra are denoted by solid lines
and the dashed lines denote the correspondent $h(q)$ spectrum plus/minus the standard
deviation of the mean for the correspondent value of $q$.}
\end{figure}

\newpage
\begin{figure}[t]
\includegraphics[angle=0,height=4.0cm,width=4.5cm]{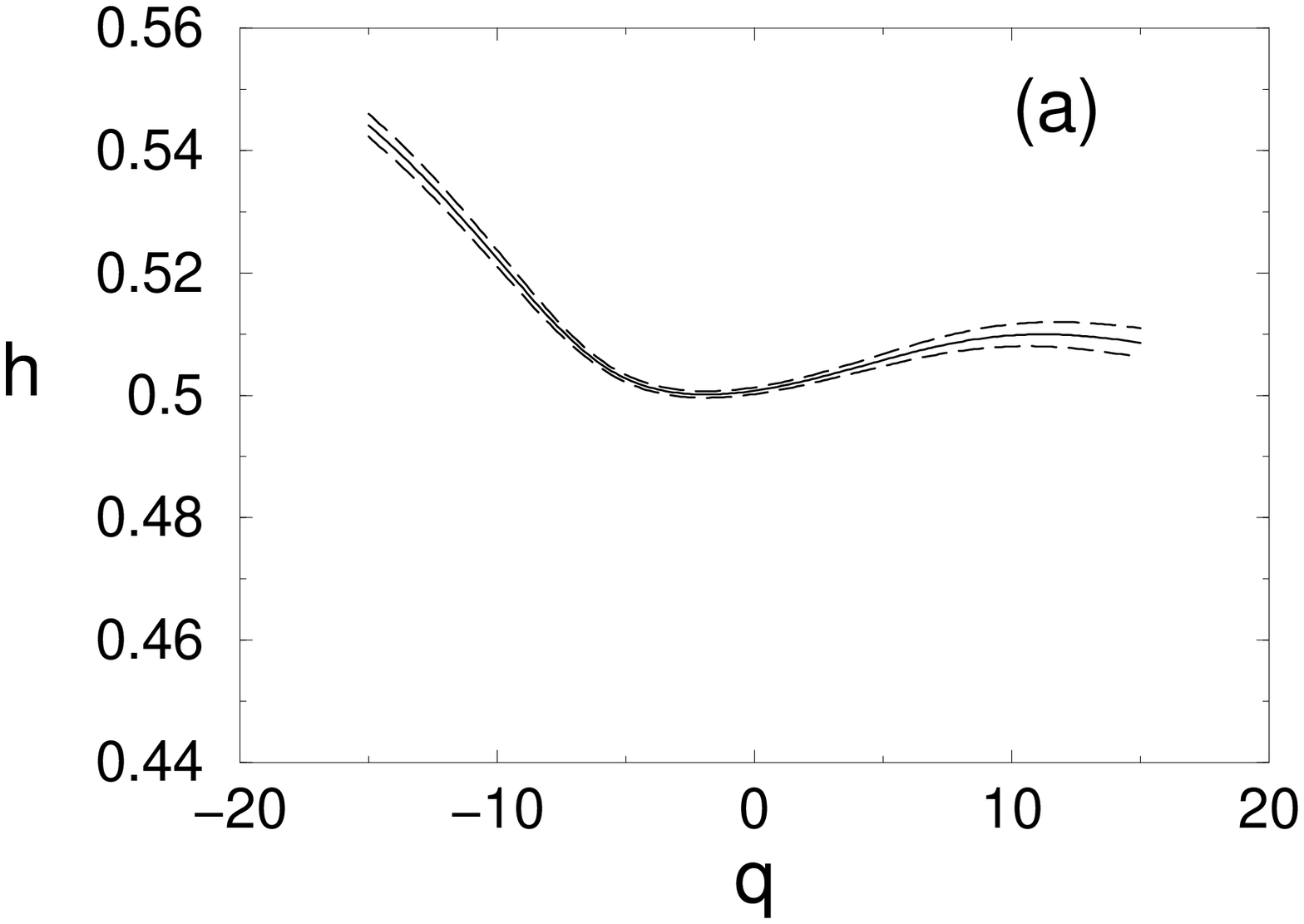} \\
\includegraphics[angle=0,height=4.0cm,width=4.5cm]{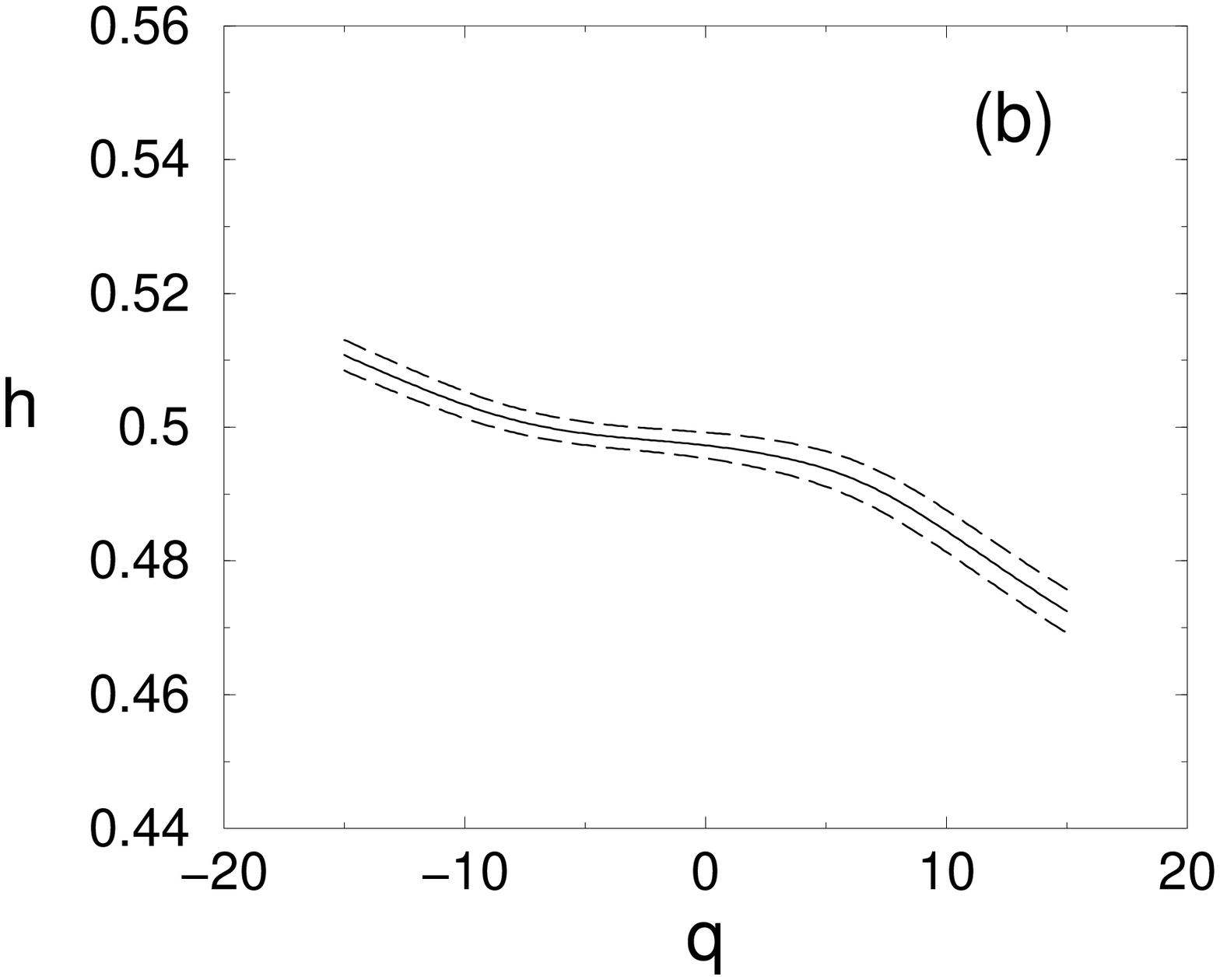} \\
\includegraphics[angle=0,height=4.0cm,width=4.5cm]{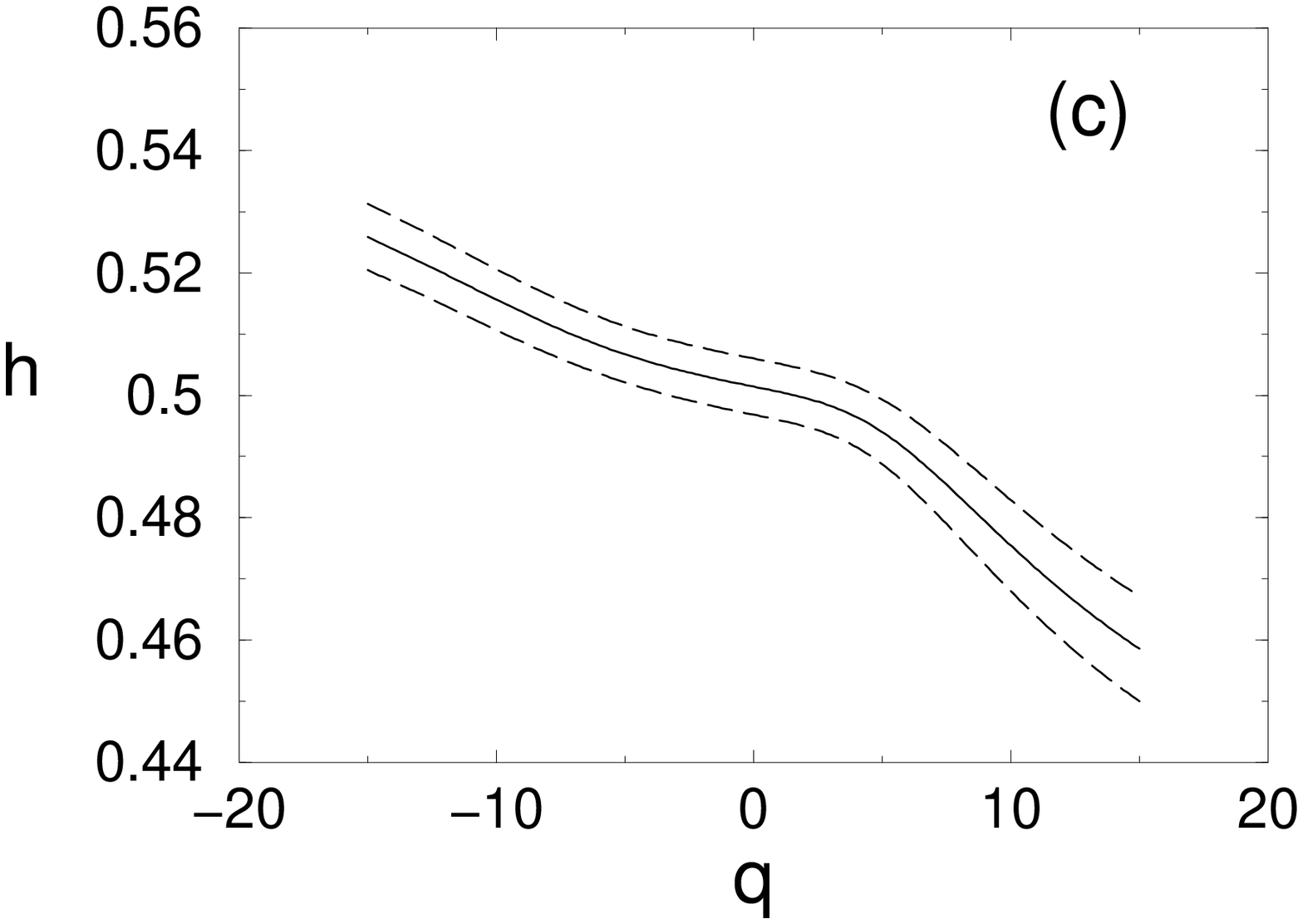}
\caption{Influence of the length of segment $s$ on the $h(q)$ spectrum of the generator
{\sf ran2}. The investigated sequences of numbers are obtained for {\sf idum=2} and have length
of $250 000$ values. The
$h(q)$ spectrum is denoted by a solid line. The two doted lines denote the $h(q)$ spectrum
plus/minus the standard deviation of the mean for the correspondent value of $q$.
Panel (a): $h(q)$ spectrum for $s$ between $s_{min}=10$ and $s_{max}=1000$. Panel (b):
$h(q)$ spectrum for $s$ between $s_{min}=100$ and $s_{max}=10000$. Panel (c):
$h(q)$ spectrum for $s$ between $s_{min}=1000$ and $s_{max}=25000$.}
\end{figure}

\begin{figure}[t]
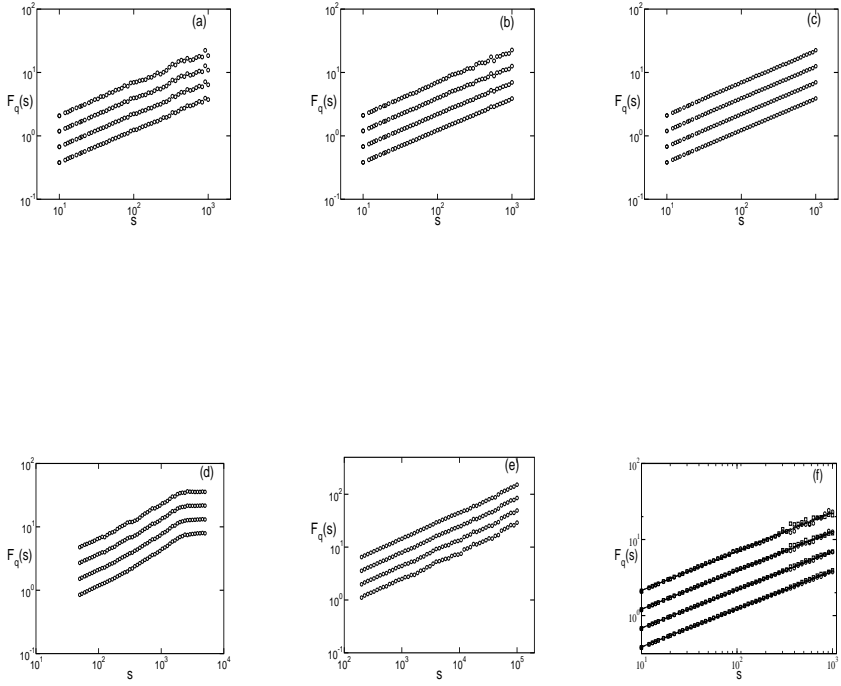

\includegraphics[angle=0,height=3.0cm,width=3cm]{figx3a.eps} \hskip.9cm
\includegraphics[angle=0,height=3.0cm,width=3cm]{figx3b.eps} \hskip.9cm
\includegraphics[angle=0,height=3.0cm,width=3cm]{figx3c.eps} \\ \vskip2.5cm   
\includegraphics[angle=0,height=3.0cm,width=3cm]{figx3d.eps} \hskip.9cm
\includegraphics[angle=0,height=3.0cm,width=3cm]{figx3e.eps} \hskip.9cm 
\includegraphics[angle=0,height=3.0cm,width=3cm]{figx3f.eps}
\caption{ Fluctuation function $F_{q} (s)$ for the pseudorandom sequences from {\sf ran2} and
{\sf ran0}. From bottom to the top at each panel the lines are for $q=2,4,6,8$. Panels (a)-(c):
Results for {\sf ran2}, {\sf idum=2}.
Panel (a): Fluctuation functions for a  sequence of $10^{4}$ points.
The values of the H{\"o}lder exponent are: $h=0.505 \pm 0.002$ ($q=2$); $h=0.505 \pm 0.004$ ($q=4$);
$h=0.503 \pm 0.004$ ($q=6$); $h=0.499 \pm 0.005$ ($q=8$).
Panel (b): Fluctuation function  for a sequence of $10^{5}$ points.
$h=0.499 \pm 0.001$ ($q=2$); $h=0.501 \pm 0.001$ ($q=4$); $h=0.505 \pm 0.002$
($q=6$); $h=0.509 \pm 0.003$ ($q=8$).
Panel (c): Fluctuation function for a sequence of $10^{6}$ points.
$h=0.500 \pm 0.0004$ ($q=2$); $h=0.502 \pm 0.001$
($q=4$); $h=0.504 \pm 0.001$ ($q=6$); $h=0.506 \pm 0.001$ ($q=8$).
Panel (d): Fluctuation function for {\sf ran0}  with
parameters $a=343$, $q=127773$, $r=2836$, $m=aq+r=43828975$ . Period of the generator is $4880$,
{\sf idum=2}.
Panel (e): Fluctuation function for random number generator  {\sf ran0} with
parameters $a=16807$, $q=127773$, $r=2836$, $m=aq+r=2147483647$. {\sf idum=2}, sequence of $10^{6}$
values.
$h=0.511 \pm 0.003$ for $q=2$; $h=0.506 \pm 0.003$
for $q=4$; $h=0.501 \pm 0.002$ for $q=6$; $h=0.498 \pm 0.002$ for $q=8$.
$a=16807$, $q=127773$, $r=2836$, $m=aq+r=2147483647$ for {\sf ran0}.
Panel (f): 
Comparison between fluctuation functions for sequences generated by the 
generators {\sf ran0} and {\sf ran2}. Length of the sequences 
$10^{5}$ points. {\sf idum=2} for the two generators. Parameters of the
generators
$a=16807$, $q=127773$, $r=2836$, $m=aq+r=2147483647$ for {\sf ran0}.
Circles: Fluctuation functions for {\sf ran2}. Squares: Fluctuation
functions for {\sf ran0}. Solid lines: Power-law fits for the sequence
generated by {\sf ran2}. From bottom to the top: $q=2,4,6,8$.
}
\end{figure}

\newpage
\begin{table} [t]
\caption{\label{tab:table1} $\chi^{2}$ test for several pseudorandom number
generators. {\sf idum}=2 for {\sf ran0, ran2, ran3}. im=6075, ia=106, ic=1283
for {\sf qdg}. }
%\begin{ruledtabular}
\begin{tabular}{cccccc}
Number of points & {\sf ran0} & {\sf ran2} & {\sf ran3} & {\sf qdg} & {\sf G05CAF} \\
\hline
$10^{4}$ & 46.30 & 39.83 & 45.13 & 7.14 & 42.25 \\
\hline
$10^{5}$ & 46.29 & 44.38 & 57.13 & 2.35 & 41.34 \\
\hline
$10^{6}$ & 42.04 & 43.29 & 38.84 & 16.54& 59.68 \\
\hline
$10^{7}$ & 41.25 & 57.33 & 41.38 & 165.84 & 56.35 \\
\end{tabular}
%\end{ruledtabular}
\end{table}

\begin{table}[t]
\caption{\label{tab:table2} Autocorrelations $c_{\nu}$ for several
pseudorandom number generators. Length of generated sequences is $10^{5}$
values. {\sf idum}=2 for {\sf ran0, ran2, ran3}. im=6075, ia=106, ic=1283
for {\sf qdg}. }
%\begin{ruledtabular}
\begin{tabular}{cccccc}
$\nu$ & {\sf ran0} & {\sf ran2} & {\sf ran3} & {\sf qdg} & {\sf G05CAF}  \\
\hline
$10^{2}$ & -0.0061  & -0.00098 & -0.0013 & 0.0017  & 0.00086 \\
\hline
$10^{3}$ & -0.0008 & -0.0033 & -0.0076 & -0.00009 & 0.0018 \\
\hline
$10^{4}$ & -0.0029 & 0.0043 & -0.0061 & -0.000098& 0.0055 \\
\hline
$5 \cdot10^{4}$ & -0.0045 & 0.0013 & -0.0023 & -0.0048 & 0.0019 \\
\end{tabular}
%\end{ruledtabular}
\end{table}

\begin{table}[t]
\caption{\label{tab:table3} H{\"o}lder exponent $h$ for several
pseudorandom number generators. Length of the pseudorandom sequences: $10^{6}$
values. $^{*}$: initialized with {\sf G05CBF}(0). $^{**}$: initialized with
{\sf G05CBF} (2).
}
%\begin{ruledtabular}
\begin{tabular}{cccccccc}
generator &   q=-2.95 & q=-1.95 & q=-0.95 & q=0.05 & q=1.05 & q=2.05 & q=3.05  \\
\hline
{\sf ran0} (idum=10) &  $0.500 \pm 0.005$  & $0.501 \pm 0.004$  & $0.502 \pm 0.004$ & $0.503 \pm 0.004$ &$0.505 \pm 0 .003$ & $0.506 \pm 0.003$ &
$0.507 \pm 0.003$     \\
\hline
{\sf ran0} (idum=20) & $0.543 \pm 0.004$  & $0.542 \pm 0.004$  & $0.540 \pm 0.004$  & $0.539 \pm 0.004$ & $0.537 \pm 0.004$ & $0.535 \pm 0.004$ & $ 0.533 \pm 0.004$  \\
\hline
{\sf ran2} (idum=10) & $0.536 \pm 0.005$  & $0.535 \pm 0.005$  & $0.535 \pm 0.005$  & $0.535 \pm 0.005$ & $0.535 \pm 0.005$ & $0.536 \pm 0.005$ & $0.538 \pm 0.005$  \\
\hline
{\sf ran2} (idum=20) & $0.533 \pm 0.007$  & $0.534 \pm 0.006$  & $0.534 \pm 0.006$  & $0.535 \pm 0.005$ & $0.535 \pm 0.005$ & $0.536 \pm 0.005$ &$0.536 \pm 0.005$  \\
\hline
{\sf ran2} (idum=4) & $0.489 \pm 0.006$ & $0.489 \pm 0.006$ & $0.489 \pm 0.006$ & $0.489 \pm 0.005$ & $0.489 \pm 0.005$ & $0.489 \pm 0.005$ & $0.490 \pm 0.005$ \\
\hline
{\sf ran3} (idum=10)&  $0.513 \pm 0.005$ & $0.512 \pm 0.005$ & $0.510 \pm 0.004$ & $0.508 \pm 0.004$ & $0.506 \pm 0.004$ & $0.504 \pm 0.004$ & $0.500 \pm 0.004$ \\
\hline
{\sf ran3} (idum=20) & $0.470 \pm 0.004$ & $0.469 \pm 0.004$ & $0.469 \pm 0.004$ & $0.470 \pm 0.004$ & $0.470 \pm 0.004$ & $0.470 \pm 0.004$ & $0.471 \pm 0.004$ \\
\hline
{\sf ran3} (idum=2) & $0.499 \pm 0.005$ & $0.499 \pm 0.005$ & $0.500 \pm 0.005$ & $0.501 \pm 0.004$ & $0.502 \pm 0.004$ & $0.503 \pm 0.004$ & $0.504 \pm 0.003$ \\
\hline
{\sf G05CAF} $^{*}$ & $0.479 \pm 0.007$ & $0.479 \pm 0.006$ & $0.480 \pm 0.006$ & $0.481 \pm 0.005$ & $0.482 \pm 0.005$ & $0.483 \pm 0.005$ & $0.484 \pm 0.005$ \\
\hline
{\sf G05CAF} $^{**}$ & $0.494 \pm 0.005$ & $0.495 \pm 0.004$ & $0.495 \pm 0.004$ & $0.496 \pm 0.004$ & $0.497 \pm 0.004$ & $0.498 \pm 0.004$ & $0.499 \pm 0.004$ \\
\hline
        \\
\end{tabular}
%\end{ruledtabular}
\end{table}

\end{document}